\documentclass[reprint, superscriptaddress, amsmath,amssymb, aps]{revtex4-1}
\usepackage{graphics}
\usepackage{graphicx}
\usepackage{subfigure}
\usepackage{dcolumn}
\usepackage{bm}
\usepackage{hyperref}
\usepackage{color}

\begin{document}
\title{Flat bands and $Z_2$ topological phases in a non-Abelian kagome lattice}
\author{Zhenxiang Gao}
\email{zhenxiang\_gao@brown.edu}
\affiliation{Department of Physics, Brown University, Providence, Rhode Island 02912, USA}
\author{Zhihao Lan}
 \email{z.lan@ucl.ac.uk}
\affiliation{Department of Electronic and Electrical Engineering, University College London,
Torrington Place, London, WC1E 7JE, United Kingdom}

\date{\today}

\begin{abstract}
We introduce a non-Abelian kagome lattice model that has both time-reversal and inversion symmetries and study the flat band physics and topological phases of this model. Due to the coexistence of both time-reversal and inversion symmetries, the energy bands consist of three doubly degenerate bands whose energy and conditions for the presence of flat bands could be obtained analytically, allowing us to tune the flat band with respect to the other two dispersive bands from the top to the middle and then to the bottom of the three bands. We further study the gapped phases of the model and show that they belong to the same phase as the band gaps only close at discrete points of the parameter space, making any two gapped phases adiabatically connected to each other without closing the band gap. Using the Pfaffian approach based on the time-reversal symmetry and parity characterization from the inversion symmetry, we calculate the bulk topological invariants and demonstrate that the unique gapped phases belong to the $Z_2$ quantum spin Hall phase, which is further confirmed by the edge state calculations. 

\end{abstract}

\maketitle

\section{\label{sec:intro}Introduction}
Motivated by both fundamental science and technological applications, exploring novel topological states of matter has recently become one of the most exciting areas of research in the condensed-matter community~\cite{Wen17RMP}. In particular, topological insulators~\cite{Kane10RMP, Qi11RMP, Bansil16RMP, BookTIandTopoSc, ShenBookTI, AsbothBookTI} have attracted a great deal of attention due to their appealing features; for example, like a conventional insulator, these insulators have an insulating bulk band gap, but host gapless conducting states at the system edge. The edge states are topologically protected by the time-reversal symmetry and thus are robust against any disorder and perturbations that do not destroy the bulk energy gap. This remarkable phase is classified based on a $Z_2$ invariant~\cite{KaneMele05PRLz2, Ryu10NJP}, which could be related to the parity of the number of gapless edge states within the bulk gap. Any even number of edge states can be shown to be adiabatically connected to a phase with no gapless edge states, and thus is topologically trivial. In contrast, an odd number of edge states can not be connected to a trivial state as long as the band gap is not closed, and thus is topologically nontrivial.

As the helical edge states of topological insulators are protected by the time-reversal symmetry, external magnetic fields are not allowed, which otherwise would break the time-reversal symmetry. In this case, the  intrinsically allowed spin-orbit coupling plays a crucial role in driving the system into the topological phases. For example, in the original Kane-Mele model \cite{KaneMele05PRL_QSH}, spin-orbit coupling between next-nearest-neighbors is essential to achieve the quantum spin Hall phase. Up to now, topological insulators have been studied in a range of two-dimensional lattice models, such as, honeycomb lattices \cite{KaneMele05PRL_QSH}, edge-centered honeycomb lattices \cite{Lan12PRB}, decorated honeycomb lattices \cite{Fiete10PRB_decorateH}, Lieb \cite{Weeks10PRB_Lieb, Goldman11PRA_Lieb, fengliu_nanolett20} or extended Lieb lattices \cite{Pal19PRB_Lieb}, kagome lattices \cite{Guo09PRB_kagome, wuming_pra10, Bolens19PRB_kagome}, ruby lattices \cite{Fiete11PRB_ruby}, and square-octagon lattices \cite{Fiete10PRB_square_octagon}. 

In contrast, recent advances in engineered quantum systems, especially cold quantum gases, allow for the creation of more complicated lattices, e.g.,  non-Abelian optical lattices, where the hopping of a multicomponent atomic gas trapped in the lattice could be tailored by a special laser configuration such that the behavior of the gas mimics particles subjected to a non-Abelian gauge potential \cite{Dalibard11RMP, Goldman14RPP, Cooper19RMP}. For example, non-Abelian lattice models have been studied in square lattices \cite{Goldman09PRL, Lan11PRB, Burrello13PRA}, honeycomb lattices \cite{Bermudez10NJP, Sun13SR}, and square-octagon lattices \cite{Fiete10PRB_square_octagon}, where very rich topological phases and phase transitions have been identified. To the best of our knowledge, kagome lattice, which has the hexagonal lattice symmetry and whose unit cell contains three sublattice sites, and its topological property have not been studied in the non-Abelian framework. For an SU(2) non-Abelian gauge potential, the generators of the Lie group, i.e., the three Pauli matrices, can naturally be associated with the three hopping directions of the kagome lattice, thus it would be interesting to see what physics one can get for such a non-Abelian generalization of the kagome lattice.

In this paper, we introduce and study a non-Abelian kagome lattice model that has both time-reversal and inversion symmetries. The model shows interesting flat bands and topological phases. Flat bands have attracted a great deal of attention recently due to a variety of interesting phenomena they can provide \cite{Ohgushi_prb00, Green10PRB_Flat, Tang11PRL_Flat, Sun11PRL_Flat, Neupert11PRL_Flat, pal_fractal_prb18, Pal18PRB_Flat, Montambaux18PRL_Flat, Jiang19PRB_Flat, fengliu_prb19, Mizoguchi19PRB_Flat, Lim20PRB_Flat, dashuai_arxiv20} and have been studied experimentally in the kagome magnet $\mathrm{Co_3Sn_2S_2}$ \cite{jiaxin_np19} and the kagome metal CoSn \cite{liu_nc20, jiaxin_nc20, Kang_nc20}. In our model, the energy bands and conditions for the presence of flat bands could be obtained analytically, allowing an apparent understanding about the flat band physics. Moreover, the gapped phases of our model all connect to each other due to the closure of a band gap occurring only at discrete points of the parameter space, i.e., they belong to the same phase. Using different techniques, we demonstrate that this unique gapped phase is a $Z_2$ topological phase.  

We would like to note that flat bands and topological phases have been studied  previously in kagome lattices \cite{Green10PRB_Flat, Guo09PRB_kagome, wuming_pra10, Bolens19PRB_kagome}. However, the main mechanism to induce topological phases in these works is spin-orbit coupling, which appears in certain solid-state materials. In the present work, the mechanism we consider is a non-Abelian gauge potential, which does not exist naturally in solid-state materials, but could be created in engineered quantum systems, such as cold atomic gases trapped in optical lattices. This setup results in analytically solvable energy band structures and interesting physics, e.g., the existence of a unique topological gapped phase in the whole parameter space. 

The paper is organized as follows. In Sec.\ref{sec:model} we present the model and discuss its symmetry properties. Flat band physics is then studied in Sec.\ref{sec:Flat band}. In Sec. \ref{sec:Z_2 topology} we investigate the topological features of the gapped phases In Sec. \ref{sec:conclusion} we summarize and provide an outlook for future work. 
        
   
\section{\label{sec:model}Model}

\begin{figure}
\includegraphics[width=\columnwidth]{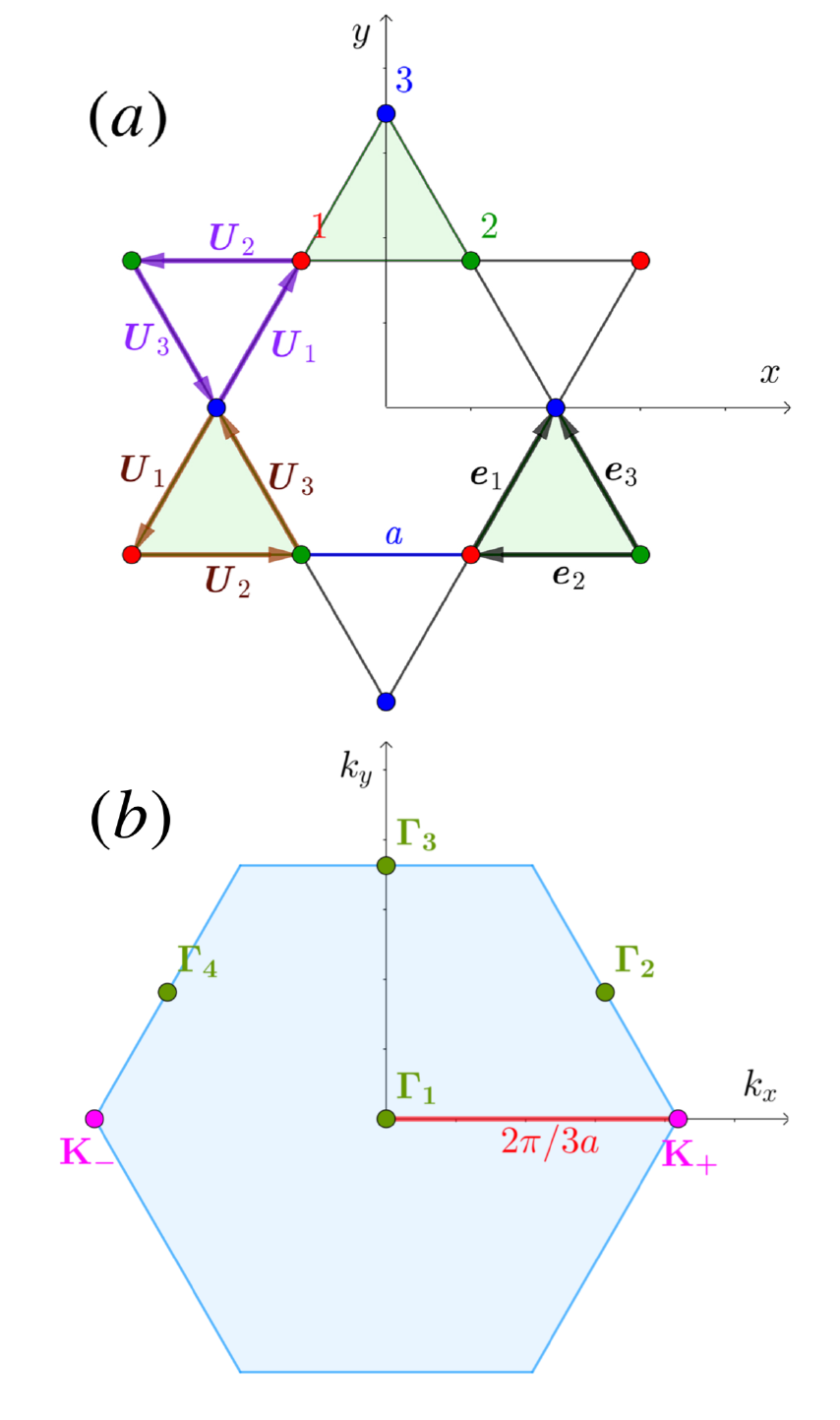} 
\caption{(a) Non-Abelian kagome lattice considered in this work. Each unit cell consists of three sublattice sites, marked as sites 1 - 3. The nearest neighbor hopping terms in the internal space are denoted by $U_1, U_2$, and $U_3$ respectively. Note that this hopping pattern preserves the inversion symmetry of the system. Here $\boldsymbol{\vec{e_1}}, \boldsymbol{\vec{e_2}}$, and $\boldsymbol{\vec{e_3}}$ denote the hopping vectors among the three sublattice sites within the unit cell, from which one can define the two lattice vectors as $2\boldsymbol{\vec{e_1}}$ and -$2\boldsymbol{\vec{e_2}}$. 
(b) First Brillouin zone of the kagome lattice, which is a hexagon with the length of its side equal to $2\pi/3a$. Here $K_\pm$ mark the two inequivalent vertices and $\Gamma_1, \Gamma_2, \Gamma_3,$ and $ \Gamma_4$ are the time-reversal invariant momenta. }
\label{fig:fig1}
\end{figure}

We consider a non-Abelian kagome lattice (see Fig. \ref{fig:fig1}), where the nearest-neighbor hopping terms of pseudospin-1/2 particles trapped in the lattice are modified according to the underlying non-Abelian gauge potential. The Hamiltonian of this system can be written as \cite{Goldman09PRL, Lan11PRB, Fiete10PRB_square_octagon},

\begin{equation} \label{Eq1}
    H=t\sum_{\langle i\tau, j \tau^{'}\rangle}[U_{ij}]_{\tau \tau^{'}}c_{i\tau}^\dagger c_{j\tau^{'}} +H.c.
\end{equation}
where the unitary matrices, $U_1, U_2$,  and $U_3$, encode the information of the non-Abelian gauge potential that the kagome lattice is subjected to.  For pseudospin-1/2 particles, a natural choice for $U_1, U_2$, and $U_3$ would be the two-dimensional representation of the $SU(2)$ Lie group, 
\begin{gather}
U_1=e^{i\alpha\sigma_1}, U_2=e^{i\beta\sigma_2}, U_3=e^{i\gamma\sigma_3}
\end{gather}
where $\sigma_1, \sigma_2,\sigma_3$ are the Pauli matrices and $\alpha, \beta, \gamma$ are parameters related to the gauge fluxes. In a real experimental implementation, e.g., cold atoms trapped in an optical lattice, the gauge flux parameters $\alpha, \beta, \gamma$ could be tuned by the amplitudes and phases of the dressing lasers that create the artificial gauge potential in the manifold of the atomic internal states (for details, see \cite{Goldman11PRA_Lieb, Lan11PRB, MazzaNJP12}).  The Hamiltonian (\ref{Eq1}) in the momentum space after a Fourier transform can be written in the form, 

\begin{equation}
    H(\boldsymbol{\vec{k}})=t\sum_{\boldsymbol{\vec{k}}}\left( \begin{array}{ccc} c_{1k}^\dagger & c_{2k}^\dagger & c_{3k}^\dagger \end{array} \right) h(\boldsymbol{\vec{k}}) \left( \begin{array}{c}
    c_{1k} \\ c_{2k} \\ c_{3k} \end{array} \right)
\end{equation}
where $c_{ik}\equiv(c_{ik\uparrow}, c_{ik\downarrow})$ with $i \in \{1,2,3\}$, i.e., the three sublattice sites in each unit cell (see Fig. \ref{fig:fig1}) and the $ 6\times 6$ Bloch Hamiltonian $h(\boldsymbol{\vec{k}})$ is given by

\begin{equation}\label{BlochH}
    h(\boldsymbol{\vec{k}})=\left( \begin{array}{ccc}
    0 & U_2^\dagger+U_2^\dagger e^{ik_2} & U_1+U_1 e^{-ik_1}\\ U_2+U_2 e^{-ik_2} & 0 & U_3^\dagger+U_3^\dagger e^{-ik_3}\\U_1^\dagger+U_1^\dagger e^{ik_1} & U_3+U_3 e^{ik_3} & 0 
\end{array} \right)
\end{equation}
where $k_\alpha\equiv \boldsymbol{\vec{k}}\cdot 2\boldsymbol{\vec{e_\alpha}} $ with $\alpha$=1,2, 3 and $\boldsymbol{\vec{e_1}}=a(1/2,\sqrt{3}/2)$, $\boldsymbol{\vec{e_2}}=a(-1,0)$ and $\boldsymbol{\vec{e_3}}=a(-1/2,\sqrt{3}/2)$, as is shown in Fig. \ref{fig:fig1}(a). In the following, we set $t=1$ and $a=1/2$ for simplicity.

Our system enjoys both the time-reversal and inversion symmetries \cite{Chiu16RMP}, characterized by 

\begin{eqnarray}
    \mathcal{T}^{-1}h (\boldsymbol{\vec{k}})\mathcal{T}&=&h(-\boldsymbol{\vec{k}}) \\
    \mathcal{P}^{-1}h(\boldsymbol{\vec{k}})\mathcal{P}&=&h(-\boldsymbol{\vec{k}})
\end{eqnarray}
where $\mathcal{T}=UK$ is the time-reversal operator with $K$ the complex conjugate operator which takes any complex number into its complex conjugate and 

\begin{equation}
    U=I_3 \otimes i\sigma_2
\end{equation}
with $I_3$ the $3\times 3$ identity matrix. Meanwhile, the inversion operator $\mathcal{P}$ (taking site 1 of the unit cell as the inversion center) is given by

\begin{eqnarray}
      \mathcal{P} & = & \left(\begin{array}{ccc}
    1 & 0 & 0\\0 & \exp(-ik_2) & 0\\0 & 0 &\exp(ik_1)
\end{array}\right)  \otimes  \left(\begin{array}{cc} 1 & 0\\0 & 1 \end{array} \right)
\end{eqnarray}
Note that while the time reversal operation flips the direction of the spin, the spin is unchanged by the inversion as the spin is a pseudovector. Furthermore, in the presence of both time-reversal and inversion symmetries, Bloch states form Kramers doublets at every $\bf{k}$ point in the Brillouin zone, i.e., the energy bands of the Bloch Hamiltonian (\ref{BlochH}) form three doubly degenerate bands. 


\section{\label{sec:Flat band} Flat bands }

The non-Abelian kagome lattice model hosts a rich phenomenon of flat bands, whose conditions of existence could be obtained analytically. Moreover, we will show that the location of the flat band could be tuned from the top to the middle and further to the bottom of the three bands.

\subsection{Conditions for the existence of flat bands}

The energy bands and corresponding eigenstates of our model could  be determined by the equation $h(\boldsymbol{\vec{k}}) |\Psi_i\rangle=E_i(\boldsymbol{\vec{k}}) |\Psi_i\rangle$, $i=1,2,\ldots,6$. For the Bloch Hamiltonian (\ref{BlochH}), the energy bands can be solved analytically. It is straightforward to show that the characteristic equation for the energy $E$ could be obtained as 
\begin{gather}\label{flatC}
(E^3-4AE-16B)^2=0
\end{gather}
where $A=\cos^2 k_1+\cos^2 k_2+\cos^2 k_3$, and $B=\Xi(\alpha,\beta,\gamma)D$ with $D=\cos k_1\cos k_2\cos k_3$, and $\Xi(\alpha,\beta,\gamma)=\cos\alpha\cos\beta\cos\gamma-\sin\alpha\sin\beta\sin\gamma$. It is now apparent that the energy bands indeed form three degenerate pairs due to the coexistence of time-reversal and inversion symmetries of the model as we described above. The three energy bands are determined by a cubic equation, whose solutions could then be obtained analytically.  

The conditions for the existence of flat bands could be derived as follows. First, using the definitions of $k_1, k_2$, and $k_3$, i.e., 

\begin{eqnarray}
    k_1 = \boldsymbol{\vec{k}}\cdot\boldsymbol{2\vec{e_1}} & = & \frac{1}{2}k_x+\frac{\sqrt{3}}{2}k_y \nonumber  \\ 
    k_2 = \boldsymbol{\vec{k}}\cdot\boldsymbol{2\vec{e_2}} & = &  -k_x \\
    k_3 = \boldsymbol{\vec{k}}\cdot\boldsymbol{2\vec{e_3}} & = & - \frac{1}{2}k_x+\frac{\sqrt{3}}{2}k_y \nonumber
\end{eqnarray} 
it is straightforward to show that 
\begin{eqnarray}
    A & = & 1+\cos(\sqrt{3}k_y)\cos (k_x)+\cos^2 (k_x)\nonumber\\
    D & = & \frac{1}{2}\cos(\sqrt{3}k_y)\cos (k_x)+\frac{1}{2}\cos^2 (k_x) 
\end{eqnarray}
 from which one can get $A=2D+1$. Then the eigenequation for the three degenerate energy bands becomes

\begin{equation}
    E^3-4(2D+1)E-16D\Xi=0
\end{equation}
or recast it in another form
\begin{equation}
\label{flt band eigval}
    E^3-4E=8(E+2\Xi)D
\end{equation}

\begin{figure}
\includegraphics[width=\columnwidth]{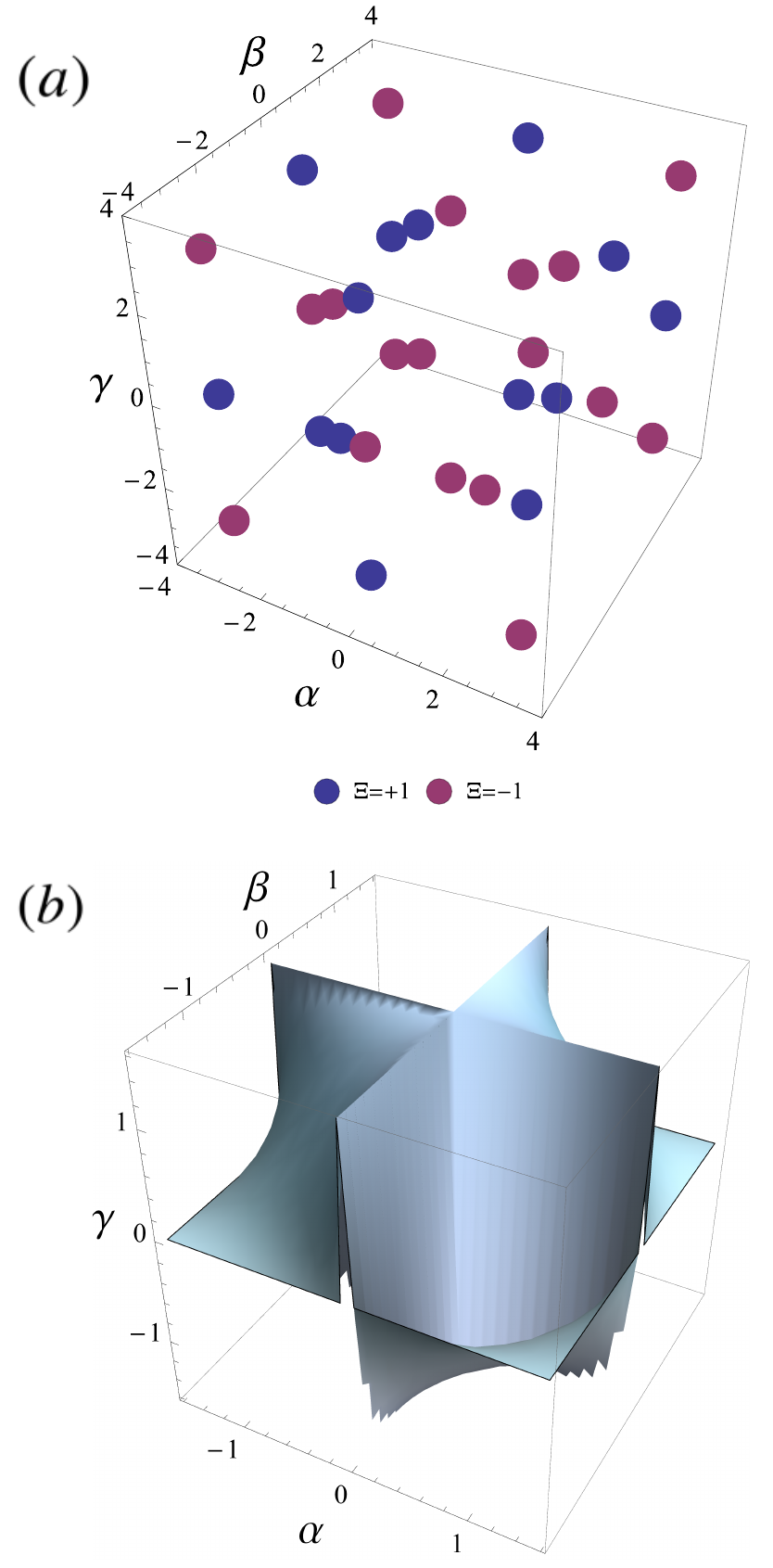} 
\caption{Constraint on $\Xi(\alpha,\beta,\gamma)$ for the existence of flat bands. (a)  $\Xi(\alpha,\beta,\gamma)=\pm1$, under which the solutions consist of discrete points, with blue points for $\Xi=+1$ and purple points for $\Xi=-1$. These discrete points correspond to the sets of gauge flux parameters giving a flat band on top (purple points) or bottom (blue points) and the system is in the gapless phase at these discrete points and in the gapped phase apart from these discrete points. (b)  $\Xi(\alpha,\beta,\gamma)=0$, under which the solutions form a surface. A point on this surface corresponds to a set of gauge flux parameters giving a flat band in the middle.}
\label{fig:fig2}
\end{figure}

\begin{figure}
\includegraphics[width=\columnwidth]{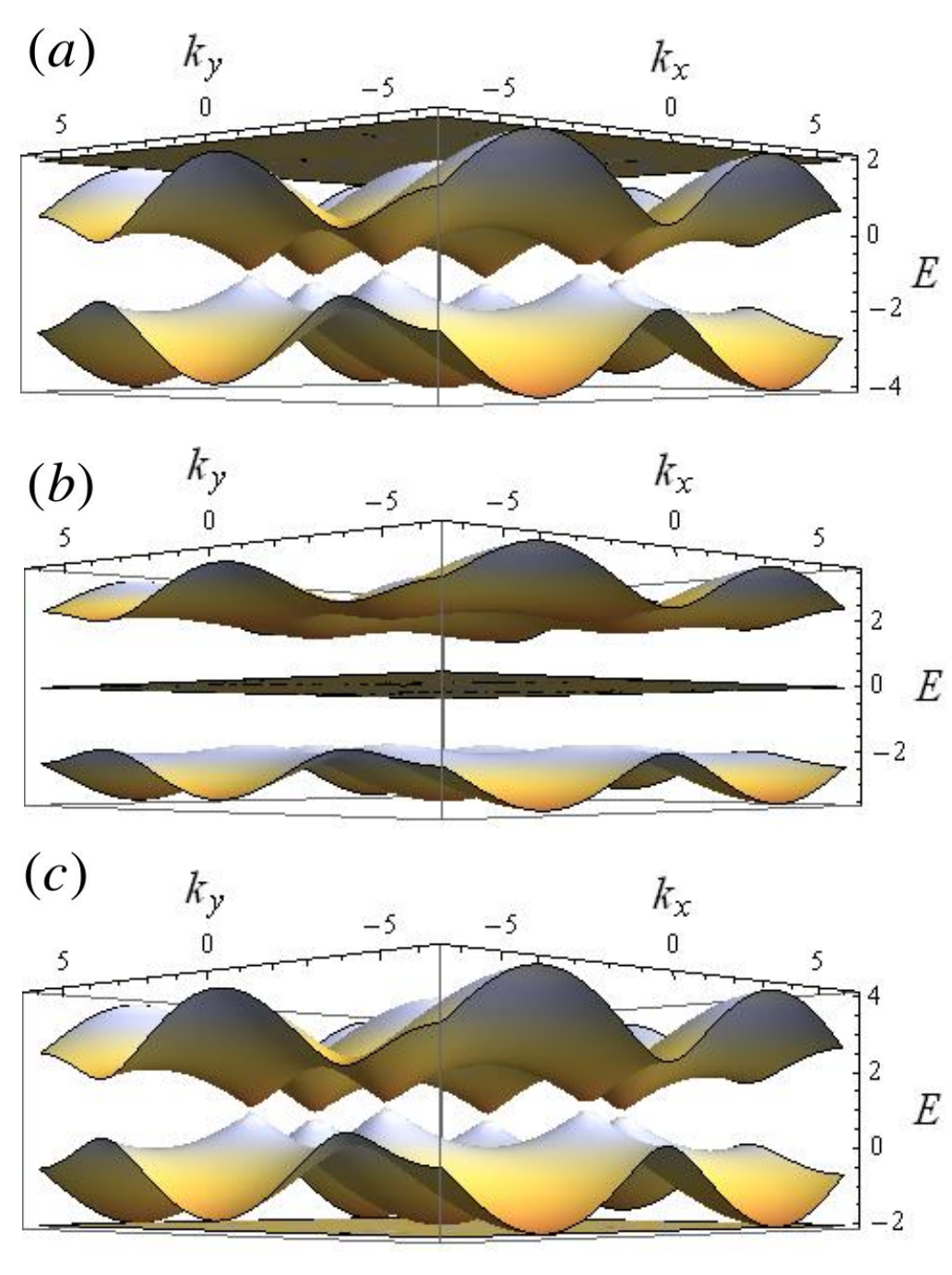} 
\caption{The location of the flat band can be tuned to: (a) the top, when $\Xi(\alpha,\beta,\gamma)=-1$; (b) the middle, when $\Xi(\alpha,\beta,\gamma)=0$ and (c) the bottom, when $\Xi(\alpha,\beta,\gamma)=1$.}
\label{fig:fig3}
\end{figure}

\noindent
For a flat band, whose energy does not depend on momentum $\bf{k}$,  one can obtain $(E=-2\Xi)$ as $D$ depends on $\bf{k}$ and is dispersive. As such one can get $E^3-4E=0$, from which one can readily obtain the energy of the three flat bands, i.e., $E=0$ and $E=\pm2$. Then from the flat band condition $E=-2\Xi$, one has $\Xi=0$ and $\Xi=\pm1$. These solutions are presented in Fig. (\ref{fig:fig2}). 

The eigenstates corresponding to the flat bands can also be obtained. For $\Xi=1$ at $\alpha=\beta=\gamma=0$, one can find
\begin{gather}
\psi^{E=-2}_1=[0,\frac{1-e^{-ik_3}}{e^{-ik_2}-1},0, \frac{e^{-ik_3}-e^{-ik_2}}{e^{-ik_2}-1},0, 1 ]^T   \\
\psi^{E=-2}_2=[\frac{1-e^{-ik_3}}{e^{-ik_2}-1},0,\frac{e^{-ik_3}-e^{-ik_2}}{e^{-ik_2}-1}, 0,1,0 ]^T  
\end{gather}
For $\Xi=0$ at $\alpha=\beta=\pi/2,\gamma=0$, one can find
\begin{gather}
\psi^{E=0}_1=[\frac{1+e^{-ik_3}}{1+e^{-ik_2}}, 0,0, \frac{i+ie^{-ik_1}}{1+e^{ik_2}},0, 1]^T \\
\psi^{E=0}_2=[0,\frac{1+e^{-ik_3}}{1+e^{-ik_2}},  \frac{i+ie^{-ik_1}}{1+e^{ik_2}},0, -1,0]^T 
\end{gather}
For $\Xi=-1$ at $\alpha=\beta=\gamma=\pi/2$, one can find
\begin{gather}
\psi^{E=2}_1=[\frac{ie^{-ik_3}-i}{e^{-ik_2}-1},0,0, \frac{ie^{-ik_2}-ie^{-ik_3}}{e^{-ik_2}-1},0, 1 ]^T   \\
\psi^{E=2}_2=[0,\frac{ie^{-ik_3}-i}{e^{-ik_2}-1},\frac{ie^{-ik_3}-ie^{-ik_2}}{e^{-ik_2}-1}, 0,1,0 ]^T  
\end{gather}

Usually, for a completely flat band in a lattice model, one can always find a compact localized state distribution. The above results show that 
the localization happens in the internal spin space due to the effect of the non-Abelian gauge potential.


\subsection{Location-tuable flat bands}

The above derived flat band conditions,  $\Xi=0$ and $\Xi=\pm1$, also determine the structure of the three doubly degenerate bands given by Eq. (\ref{flatC}), which are presented in Fig.(\ref{fig:fig3}). It is evident that the location of the flat band with respect to the two dispersive bands could be tuned from the top ($E=+2$) to the middle ($E=0$) and further to the bottom ($E=-2$). The band structure for $\Xi=1$ is the same as that of the normal kagome lattice ~\cite{Guo09PRB_kagome}. In fact, when $\alpha=\beta=\gamma=0$, $\Xi(\alpha,\beta,\gamma)=1$ could be trivially satisfied as in this case $U_1$, $U_2$ and $U_3$ reduce to identity matrices and consequently, the non-Abelian kagome lattice will reduce to two uncoupled copies of the normal kagome lattice. 

It would be interesting to compare the flat bands in the current non-Abelian kagome lattice model and other kagome models subjected to the Abelian gauge potential \cite{Ohgushi_prb00, Green10PRB_Flat, kagomeSKY}. For example, the authors of Ref.~\cite{Green10PRB_Flat} have shown the existence of isolated flat bands and spin-1 conical bands with a flat band located at $E=0$ employing staggered flux phases $\psi_+$ and $\psi_-$ on the up and down triangles of the kagome lattice. A recent work ~\cite{kagomeSKY} considering the same flux $\psi$ on the up and down triangles of the kagome lattice has shown that the flat band in the model can also be tuned from the top to the middle and further to the bottom when changing the flux $\psi$.  The present results on the conditions for the existence of flat bands and their tunability certainly will enrich the flat band physics in kagome lattices from the Abelian to the non-Abelian regime.


\section{\label{sec:Z_2 topology}$Z_2$ topological phases}

Apart from the gapless phase when the flat band is at the top or the bottom, which corresponds to a Dirac semimetallic phase, where the topological properties of the Dirac cones in Fig. (\ref{fig:fig3}) are well known \cite{BookTIandTopoSc, Guo09PRB_kagome}, such as the existence of the Berry phase of $\pm\pi$ around the Dirac points $K/K'$, our model also hosts gapped insulating phases, e.g., see Fig.\ref{fig:fig3}(b), thus it is a natural question whether such phases are topological. In the following, we first present the gap phase diagram when tuning $\alpha, \beta, \gamma$ and then we show using different techniques that the gapped phases are $Z_2$ topological phases. 

\subsection{energy gaps}

\begin{figure}
\includegraphics[width=\columnwidth]{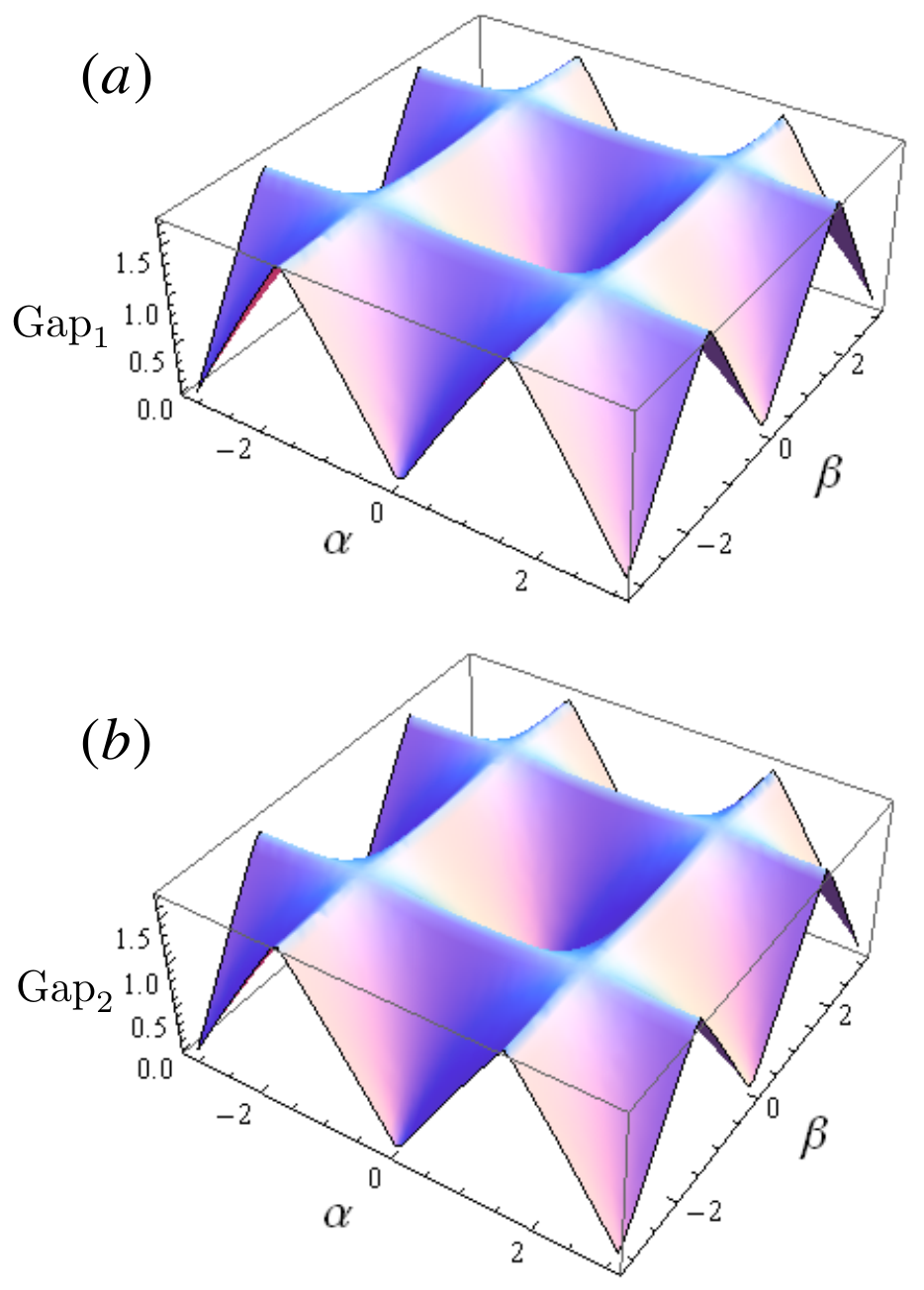} 
\caption{Sizes of the two band gaps as functions of $\alpha, \beta$ for $\gamma=0$.  One can see only at some discrete points in the $(\alpha,\beta)$ plane that the band gap closes and consequently, all the gapped phases are adiabatically connected, i.e., they belong to the same phase.  }
\label{fig:fig4}
\end{figure}

We begin by studying the gap phase diagrams when tuning $\alpha, \beta, \gamma$.  For simplicity,  we fix $\gamma=0$ and vary $\alpha, \beta$, with the sizes of the two band gaps presented in Fig. (\ref{fig:fig4}). First, one can see that the two gaps have the same behavior as functions of $\alpha, \beta$. Second, the two gaps only close at some discrete points in the  $(\alpha,\beta)$ plane, e.g., $(0,0)$. Moreover, as the gauge flux parameters $\alpha, \beta, \gamma$ group into a single parameter $\Xi$ that determines the band structure, one can easily show from Eq. (\ref{flatC}) that the band gap closes only at $\Xi=\pm1$. In other words,  Fig. \ref{fig:fig2} (a) can be taken as the phase diagram of our model, where the discrete points correspond to the gapless semimetallic phases whereas other regions correspond to the gapped insulating phases. As a consequence, one can make a remarkable statement that all the gapped phases in Fig. \ref{fig:fig2} (a) belong to the same phase as any two gapped phases could be connected adiabatically by tuning ($\alpha, \beta, \gamma$) without closing the band gap. In the following, we will try to demonstrate using different techniques that this unique gapped phase is a $Z_2$ topological phase by focusing on $\alpha=\beta=\pi/2$ and $\gamma=0$, whose band structure is shown in Fig. \ref{fig:fig3} (b).

\subsection{Pfaffian characterization of the gapped phases from time-reversal symmetry}

\begin{figure}
\includegraphics[width=\columnwidth]{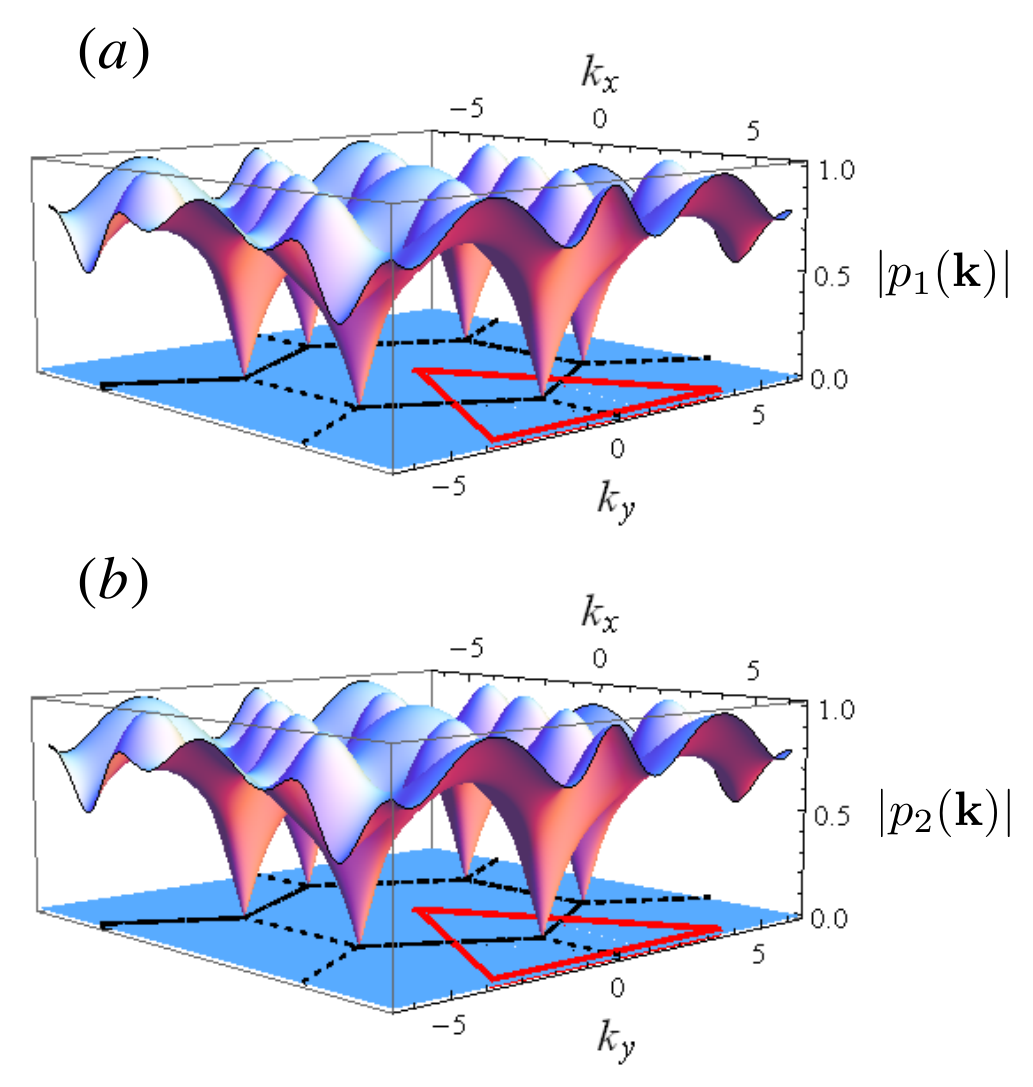} 
\caption{Modulus of the Pfaffian ($P(\boldsymbol{\vec{k}})$ as in Eq. \ref{eqn:P(k)}) with the first Brillouin zone shown in the $(k_x,k_y)$ plane for (a) the lower band gap and (b) the higher band gap at $\alpha=\pi/2$, $\beta=\pi/2$ and $\gamma=0$.  It can be seen that there are six zeros of the Pfaffian located at $K_{\pm}$ and thus there is only one zero of the Pfaffian in half of the Brillouin zone (red triangle), indicating $Z_2=1$.} 
\label{fig:fig5}
\end{figure}

We note that the Berry curvature $\mathcal{F}({\bf k})$ of the bands of our model is zero due to the simultaneous presence of time-reversal and inversion symmetries in our model. This is because under time-reversal, the Berry curvature is odd, i.e., $\mathcal{F}({\bf -k})=-\mathcal{F}({\bf k})$ and under inversion, it is even, i.e., $\mathcal{F}({\bf -k})=+\mathcal{F}({\bf k})$ and as a result $\mathcal{F}({\bf k})=0$ \cite{fu2007PRB}. So to characterize the topology, we first employ a method based on the Pfaffian originally proposed by Kane and Mele \cite{KaneMele05PRLz2} to classify the gapped phase. For a system satisfying the time reversal symmetry, one can define a matrix  $m_{ij}(\boldsymbol{\vec{k}})=\langle u_i(\boldsymbol{\vec{k}})|\mathcal{T}|u_j(\boldsymbol{\vec{k}})\rangle$, where $\mathcal{T}$ is the time-reversal operator, $|u_\alpha(\boldsymbol{\vec{k}})\rangle$ is the periodic part of the Bloch eigenstate and $i,j=1 \cdots N$ with $N$ the number of occupied bands. It can be proved that  $m_{ij}(\boldsymbol{\vec{k}})$ is skew-symmetric \cite{BookTIandTopoSc}, i.e., $m_{ij}(\boldsymbol{\vec{k}})=-m_{ji}(\boldsymbol{\vec{k}})$ and as such we can define

\begin{equation}
\label{eqn:P(k)}
    P(\boldsymbol{\vec{k}})=\textrm{Pf}\,[\langle u_i(\boldsymbol{\vec{k}})|T|u_j(\boldsymbol{\vec{k}})\rangle]
\end{equation}
where Pf[$A$] means the Pfaffian of a skew-symmetric matrix $A$, i.e., Pf[$A$]=$\sqrt{\det(A)}$. Then, the $Z_2$ topological invariant can be determined by the zeros of $P(\boldsymbol{\vec{k}})$~\cite{KaneMele05PRLz2}
\begin{equation}
\label{eqn:Z2 by integrating}
    Z_2=\frac{1}{2\pi i}\oint_C d\boldsymbol{\vec{k}} \cdot  \nabla_{\boldsymbol{\vec{k}}} \log[P(\boldsymbol{\vec{k}})] \hspace{0.5cm} \text{mod} 2
\end{equation}    
where $C$ is the path that surrounds half of the first Brillouin zone (i.e., the red lines) as shown in Fig. \ref{fig:fig5}.  

If there is only one zero of the Pfaffian in half of the Brillouin zone, it is stable globally. The reason is that, similar to the case of the Dirac node in graphene, the zero of the Pfaffian has a vorticity and since $P(-\boldsymbol{\vec{k}})$ is related to $P^*(\boldsymbol{\vec{k}})$, the phases of the Pfaffian close to $-\boldsymbol{\vec{k}}$ and $\boldsymbol{\vec{k}}$ have opposite order. If there is one zero in half of the Brillouin zone, the only place where the zeros can get annihilated is a T-invariant point. However, one can show that the Pfaffian at the T-invariant point has unit modulus \cite{BookTIandTopoSc}, thus they can not annihilate at the T-invariant point. 

The absolute values of the Pfaffian of the two gaps are presented in Fig.\ref{fig:fig5} along with the first Brillouin zone located in the $(k_x, k_y)$ plane, where we also show the integration path of Eq. (\ref{eqn:Z2 by integrating}) along boundaries of half the Brillouin zone (red triangle). It can be seen that  the zeros are located at $K_{\pm}$ and inside half of the Brillouin zone, there is only one zero, thus $Z_2=1$, indicating that the gapped phase is a $Z_2$ topological phase. We also check that for other gapped phases in Fig.\ref{fig:fig2} (a), the Pfaffian has a similar distribution in the first Brillouin zone, i.e., with six zeros located at $K_{\pm}$, consistent with our statement that all the gapped phases belong to the same topological phase. 

\subsection{Parity characterization of the gapped phases from inversion symmetry}

When the system has inversion symmetry, the evaluation of the $Z_2$ invariants could  be greatly simplified. In particular, Fu and Kane showed that the $Z_2$ invariants in this case can be determined from the parity of the occupied Bloch wave functions at the time-reversal invariant points in the Brillouin zone ~\cite{fu2007PRB}. As our non-Abelian kagome lattice model preserves the inversion symmetry, it would be interesting to see the parity effect on the $Z_2$ characterization of the gapped phases. 

In this method, as the inversion operation commutes with the Bloch Hamiltonian at the time-reversal invariant points, the parity $\xi_{2m}(\Gamma_i)=\pm 1$ of the   $(2m)$-th energy band can readily be evaluated, from which one could define the parity effect at one time-reversal invariant point for all the occupied bands, 

\begin{equation}
    \delta_i=\prod_{m=1}^{N} \xi_{2m}(\Gamma_i)
\end{equation} 
where $2N$ is the number of occupied bands. Note, that the energy band is doubly degenerate at the time-reversal invariant points due to the Kramers theorem, and here we only need to include one, i.e., $2m$, of the two degenerate partners. 

From $\delta_i$, we can obtain the value of $\nu$ taking the parity effect at all the time-reversal invariant points into account, which is a $Z_2$ quantity that distinguishes the non-trivial topological phase from the trivial phase through
\begin{equation}
    (-1)^\nu=\prod_{i} \delta_i
\end{equation}   
Here the value of $\nu$ can be 0, which implies the normal phase, or 1, which implies the $Z_2$ topological phase.

The $Z_2$ characterization of the gapped phases is shown in Table \ref{tab:z2}, from which one can see that both gaps host $\nu=1$, i.e., the gapped phases for both band gaps are $Z_2$ topological phases, consistent with the Pfaffian characterization described above. We further that check this $Z_2$ characterization is independent of $\alpha, \beta, \gamma$ as long as the gap exists.

\begin{table}
\centering
\caption{The $Z_2$ topological invariants of the gapped phases at $\alpha=\pi/2$, $\beta=\pi/2$ and $\gamma=0$, which show that both of the two energy band gaps have $\nu=1$, corresponding to a $Z_2$ topological phase.}
\begin{tabular}{|c|c|c|c|c|c|c|}
    \hline
      $2m$ & $\xi_{2m}(\Gamma_1)$ & $\xi_{2m}(\Gamma_2)$ & $\xi_{2m}(\Gamma_3)$ & $\xi_{2m}(\Gamma_4)$ & $\prod_i \xi_{2m}(\Gamma_i)$ & $\;\nu\;$ \\
    \hline
     2 & +1 & -1 & +1 & +1 & -1 & 1\\
    \hline
     4 & +1 & +1 & -1 & -1 & +1 & 1\\
    \hline
     \end{tabular}
     \label{tab:z2}
\end{table}


\subsection{Edge state characterization of the gapped phases from bulk-edge correspondence}

According to the bulk-edge correspondence principle \cite{BookTIandTopoSc, ShenBookTI, AsbothBookTI}, when the gapped bulk phase is topological, there will be edge states emerging within the topological band gap. Here, we will study the edge state, which is additional evidence that the gapped phase is topological. 

\begin{figure}
\includegraphics[width=\columnwidth]{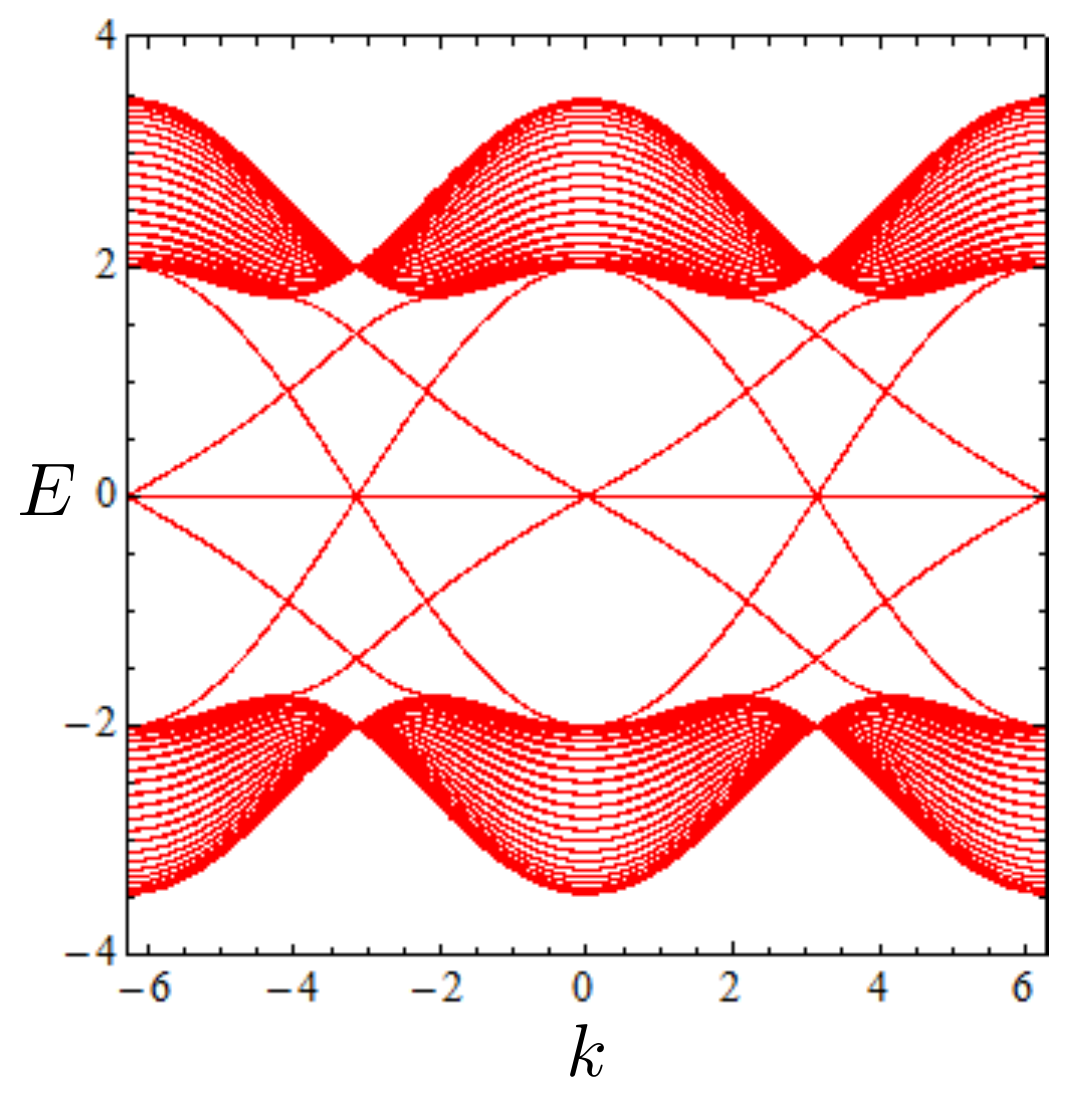} 
\caption{Band structure of a finite size non-Abelian kagome lattice which is periodic in the $x$ direction and finite in the $y$ direction $(N=20)$ at $\alpha=\pi/2$, $\beta=\pi/2$ and $\gamma=0$. The existence of spin-momentum locking edge states within both of the two band gaps indicates the quantum spin Hall nature of the two topological gaps.}
\label{fig:fig6}
\end{figure}

To study the edge states, we consider a supercell which is periodic along x but finite ($N=20$) in the y direction and present the band structure in Fig. (\ref{fig:fig6}).  Note that, to show more clearly how the edge states connect with each other across the one-dimensional Brillouin zone, we show the band structure for $k\in\{-2\pi, 2\pi\}$. As can be seen, apart from the bulk states and the flat band, which already show up in Fig. \ref{fig:fig3}(b), there are additional edge states appearing inside the two band gaps, which connect with the bulk states at the top and bottom of the band gap. These edge states can not be removed into the bulk as long as the band gap exists, indicating their topological nature. Furthermore, we can see that within each band gap, there are four edge states, two along each open edge in the y direction.  At the same edge, one edge state has a negative group velocity (i.e., the slope of the edge state  dispersion curve is negative), while the other has a positive group velocity, indicating spin-momentum locking, a hallmark for quantum spin Hall states due to the $Z_2$ topological invariant.  We also check that as long as the band gap persists, there will always be edges states within the two gaps, indicating that the gapped phases belong to the same $Z_2$ topological phase. 

\section{\label{sec:conclusion} Conclusions and Outlooks}

We have introduced a non-Abelian kagome lattice model that hosts interesting flat bands and $Z_2$ topological phases. The model lattice system has both time-reversal and inversion symmetries, resulting in three doubly degenerate bands. The energy bands of the system and the conditions for the presence of the flat bands have been obtained analytically, allowing for a transparent understanding of the flat band physics. By tuning the non-Abelian gauge flux parameters, the flat band with respect to the other two dispersive bands could be tuned from the top to the middle and further to the bottom of the three bands. Furthermore, our system also hosts gapped phases. We found that the gapped phases all connect to each other for both of the two band gaps as the gaps only close at certain discrete points in the parameter space, thus all the gapped phases belong to the same phase. Using different techniques, such as the Pfaffian approach based on  time-reversal symmetry and parity characterization from the inversion symmetry, we have demonstrated that the bulk topological invariant is equal to $Z_2$, and thus the gapped phase is a $Z_2$ topological phase. We also studied the edge states according to the bulk-edge correspondence principle and found that spin-momentum locking edge states emerge within the two band gaps, further validating the quantum spin Hall nature of these edge states. 

The fact that the gaps host a single $Z_2$ topological phase certainly is appealing as this allows for easy observation in experiments and promising applications in practice. The non-Abelian kagome lattice model introduced in this work will open up other interesting directions of investigation, such as interaction effects \cite{Rachel18RPP_review}, disorder physics \cite{kagomeSKY}, or even non-Hermitian physics \cite{Bergholtz19arXiv, Ashida20arXiv}. Furthermore, the non-Abelian gauge potentials could be implemented in such a way that the inversion symmetry of the model is broken, and the fate of flat bands and the gapped phases in this scenario is also an interesting question. One could also consider the non-Abelian kagome lattice with higher pseudospin, such as the $SU(3)$ model for pseudospin-1 particles \cite{Barnett12PRL, Bornheimer18PRA} and even the higher-order topological phases, such as the second-order corner states \cite{Parameswaran17physics, Ezawa18PRL_corner} within the present setup. Thus we believe our work will stimulate further research topics in this lattice model.



\end{document}